\documentclass[a4paper,11pt]{article}

\usepackage[letterpaper,top=3cm,bottom=3cm,left=2.5cm,right=2.5cm,marginparwidth=1.75cm]{geometry}

\usepackage[utf8]{inputenc}
\usepackage[sort&compress,numbers]{natbib}
\usepackage{enumitem}
\usepackage[table]{xcolor}
\usepackage[colorinlistoftodos]{todonotes}
\usepackage{color, soul}
\usepackage{multirow}
\usepackage{colortbl}
\usepackage{subcaption}
\usepackage{caption}
\usepackage{svg}
\usepackage{longtable}
\usepackage[colorlinks=false, hidelinks]{hyperref}

\usepackage{booktabs}
\usepackage{longtable}
\usepackage{comment}
\usepackage{adjustbox}
\usepackage{authblk}

\usepackage{tikz}
\def\checkmark{\tikz\fill[scale=0.4](0,.35) -- (.25,0) -- (1,.7) -- (.25,.15) -- cycle;} 

\title{\textbf{Software analytics for software engineering: A tertiary review}}

\author{\textbf{Muhammad Laiq\textsuperscript{a[0000--0002--5964--5554]}, Nauman bin Ali\textsuperscript{a[0000--0001--7266--5632]}, Jürgen Börstler\textsuperscript{a[0000--0003--0639--4234]}, Emelie Engström\textsuperscript{b[0000--0001--6736--9425]}}\\ 
\textsuperscript{a} Department of Software Engineering, Blekinge Institute of Technology, Sweden \\
(muhammad.laiq, nauman.ali, jurgen.borstler)@bth.se \\
\textsuperscript{b} Department Computer Science, Lund University, Lund, Sweden \\
(emelie.engstrom)@cs.lth.se}

\date{}

\begin{document}
\maketitle

\section*{Abstract}
Software analytics (SA) is frequently proposed as a tool to support practitioners in software engineering (SE) tasks. 
We have observed that several secondary studies on SA have been published.
Some of these studies have overlapping aims and some have even been published in the same calendar year. This presents an opportunity to analyze the congruence or divergence of the conclusions in these studies. Such an analysis can help identify broader generalizations beyond any of the individual secondary studies.

We identified five secondary studies on the use of SA for SE. These secondary studies cover primary research from 2000 to 2021. Despite the overlapping objectives and search time frames of these secondary studies, there is negligible overlap of primary studies between these secondary studies. Thus, each of them provides an isolated view, and together, they provide a fragmented view, i.e., there is no ``common picture'' of the area. Thus, we conclude that an overview of the literature identified by these secondary studies would be useful in providing a more comprehensive overview of the topic.

\vspace{5mm}
\noindent\textbf{Keywords:} Software engineering, Software analytics, Tertiary review, Machine learning, Data analytics, Visual analytics

\section{Introduction}

Throughout the software life cycle, a large amount of data is produced, embedded in for example, bug reports, test cases, test executions, version control systems, and source code. These data can be used to support practitioners in decision-making processes regarding their daily activities. 
However, such insights are not directly available from the data.
Software analytics (SA) has emerged as a systematic approach to utilize data embedded in software engineering artifacts to generate insights for supporting practitioners in the decision-making process \cite{zhang2013sainpractice}. SA has been proposed in various areas of software engineering (SE) to support various practitioner roles, such as testers \cite{marijan2013test}, developers \cite{rakha2018revisiting}, and managers \cite{laiq2024industrial}.

Several secondary studies (e.g., \cite{abdellatif_software_2015,anwar_towards_2017}) have been conducted to provide an overview of the use of SA for SE.
The publication years of these studies are close to each other, and they cover overlapping topics. This presents an interesting opportunity to investigate them together and compare and contrast their definition of SA, search and selection approach for identifying relevant literature, and their conclusions. By comparing the findings of these studies, a tertiary review can identify potential gaps, overlaps, or inconsistencies in the literature. In addition, it can provide a comprehensive understanding of the current state of research in this field that can guide future research.
In this report, we present a tertiary review to analyze existing secondary studies on the usage of SA for SE.

Several tertiary studies \cite{garousi2016systematic, IftikharABU24,TranUBA21,BorstlerAP23,zhou2016map,nurdiani2016impacts,pillai2017consolidating} have been conducted in SE. However, to our knowledge, there is no tertiary review focusing on the usage of SA for SE

The remainder of the report is structured as follows. In Section \ref{sec:research-method}, we describe the research method. In Section \ref{phase-1-results}, we present the results. In Section \ref{sec:discussion}, we discuss the findings. Finally, in Section \ref{sec:conclusion-future-work}, we conclude our report with future work.

\section{Research method}\label{sec:research-method}

\begin{figure}[!ht]
    \centering
    \includegraphics[width=0.7\textwidth]{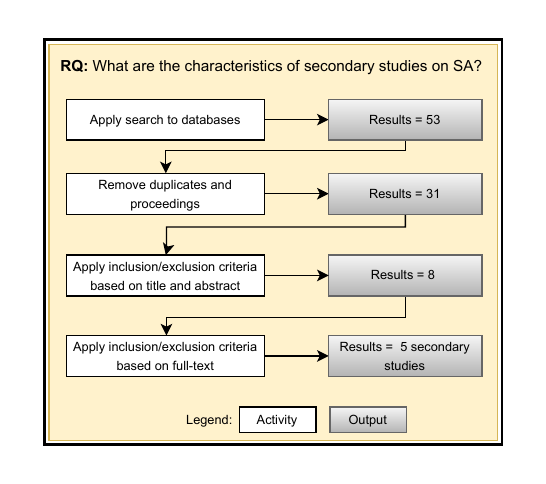}
    \caption{Overview of our search process}
    \label{fig:sms-process-phase-1}
\end{figure}

We pose and answer the research questions listed in Table \ref{tab:phase-1-rqs} following the guidelines of Kitchenham et al. \cite{kitchenham2015evidence} for conducting systematic literature reviews.
Figure~\ref{fig:sms-process-phase-1} presents an overview of our search process. 

\begin{table}[h]
    \centering
    \caption{Research questions}
    \vspace{-2mm}
    \label{tab:phase-1-rqs}
    \begin{tabular}{p{6.5cm}l}
    \toprule
     \textbf{Research question} & \textbf{Sub-research questions} 
    \\ \midrule 
     \multirow{4}{6.5cm}{RQ: What are the characteristics of secondary studies on SA?}
     & RQ1: How is SA defined? \\ 
     & RQ2: Which topics have been studied? \\
     & RQ3: What search strategies have been used? \\ 
     & RQ4: What is the overlap of primary studies? 
    \\ \bottomrule
\end{tabular}
\end{table} 

\subsection{Search process}\label{sec:search-phase-1}

Table \ref{tab:searchstring} shows our search string in selected databases to find relevant literature on SA.
Our search string is focused on secondary studies that explicitly discuss SA.
To find papers that use phrasings such as analytics for software development, or analytics for software testing, or software development analytics, we utilized the proximity operators supported by Scopus, IEEE Xplore, and Web of Science. For example, in Scopus, `software w/1 analytics' will cover the occurrence of `software' and `analytics' with at most one word between them, regardless of their order. We changed the number from 1--3. However, we could not find any additional relevant papers. Thus we decided to keep n=1.

Table \ref{tab:db-hits} shows the number of search results in each database. In the following, we summarize the results of the search in selected databases.

\begin{itemize}

\item As shown in Table \ref{tab:db-hits}, our search found 53 hits in selected databases, performed on April 2, 2023. Five secondary studies of the 53 were selected after applying the selection criteria (see Section \ref{sec:study-selection-phase-1}).

\item Of the five selected secondary studies (i.e., SS1, SS2, SS3, SS4, and SS5, see Table \ref{tab:selected-SSs-phase-1}), SS1 was found in all four databases. SS2, SS3, and SS4 were found only in Scopus and the Web of Science. SS5 was found in Scopus, Web of Science, and IEEE Xplore. This implies that we could have found all the selected secondary studies using any of the indexing databases, that is, Scopus or Web of Science.

\item Furthermore, the search in Scopus produced nine unique papers,  and the search in IEEE Xplore and Web of Science produced ten and three unique papers, respectively. However, we could not find any unique paper in ACM.

\end{itemize}

\begin{table}[ht]
    \centering
    \caption{Search strings in selected databases}
    \vspace{-2mm}
    \label{tab:searchstring}
    \begin{tabular}{p{3.4cm}p{12cm}}
    \toprule
     \textbf{Database/Search engine} & \textbf{Search string}
    \\ 
    \midrule
     Scopus &  \textit{TITLE-ABS-KEY ((software w/1 analytics )  AND  (``literature review"  OR  ``systematic review"  OR  ``systematic map"  OR  ``systematic mapping" OR  ``systematic scoping"))}
    \\ 
    \midrule
    IEEE Xplore &  Full-text search was performed using the command feature of IEEE Xplore: \textit{ ((software NEAR/1 analytics ) AND (``literature review" OR ``systematic review" OR ``systematic map" OR ``systematic mapping" OR ``systematic scoping")) }
    \\ 
    \midrule
    ACM  & Search was performed on title, abstract, and keywords and combined using the OR operator utilizing the query editor feature of ACM: ((``software analytics" OR ``software development analytics") AND (``literature review" OR ``systematic review" OR ``systematic map" OR ``systematic mapping" OR ``systematic scoping"))
    \\ 
    \midrule
    Web of Science  & Search was performed on title, abstract, and keywords and combined using the OR operator: \textit{ ((software NEAR/1 analytics) AND (``literature review" OR ``systematic review" OR ``systematic map" OR ``systematic mapping" OR ``systematic scoping")) } 
    \\ 
    \bottomrule

    \end{tabular}

\end{table}

\begin{table}[ht]
    \centering
    \caption{Database and search results}
    \vspace{-2mm}
    \label{tab:db-hits}
    \begin{tabular}{p{4cm}p{3cm}}
    \toprule
     \textbf{Database} & \textbf{Search results} 
    \\  \midrule 
    Scopus & 23
    \\  \midrule
    IEEE Xplore & 15
    \\  \midrule
    ACM & 1
    \\ \midrule
    Web of Science & 14
    \\ \midrule
    
     & \textbf{Total = 53}
    \\ \bottomrule

    \end{tabular}

\end{table}

\subsection{Study selection}\label{sec:study-selection-phase-1}

We selected papers based on title, abstract reading, and full-text reading. We performed full-text reading only when we could not decide whether to include or exclude a paper based on the title and abstract reading. The following inclusion criteria were followed for study selection:

\begin{itemize}

\item [I1:] The paper is peer-reviewed.

\item [I2:] The paper is a secondary study.

\item [I3:] The paper is positioned explicitly as an SA review, i.e., it explicitly reports that the review has been performed on SA-based work. 

\end{itemize}

Figure \ref{fig:sms-process-phase-1} shows the number of included and excluded papers at each stage. As shown in Figure \ref{fig:sms-process-phase-1}, full-text reading was performed on eight papers. Of these eight papers, five were selected, and three were excluded. Of those three excluded, two papers did not explicitly discuss SA, and one did not provide the list of selected primary studies on SA.

Two reviewers were involved in the selection process at each stage. The pilot was carried out before the actual study selection process to ensure mutual understanding. During the selection process, we had disagreements on seven papers, which we resolved through discussion and reconsideration. We discussed and agreed to include only those papers that explicitly talk about SA and an SE context.

\subsection{Data extraction}\label{sec:data-extraction-phase-1}

Table \ref{tab:dex-template-1} shows the data extraction template we used to extract data from the selected secondary studies on SA.

\begin{table}[ht]
    \centering
    \caption{Data extraction template}
    \vspace{-2mm}
    \label{tab:dex-template-1}
    \begin{tabular}{lp{14cm}}
    \toprule
     \textbf{Item} & \textbf{Description}
    \\ 
    \midrule 
    \#1 & Definition of SA 
    \\ 
    \midrule
    \#2 & Topic on which secondary study is focused, e.g., SA for agile software development
    \\ 
    \midrule
        \#3 & Study publication year
    \\ 
    \midrule
        \#4 & Search strategy used
    \\ 
    \midrule
        \#5 & Databases used for identifying literature on SA
    \\ 
    \midrule
        \#6 & Timeline covered by search
    \\ 
    \midrule
        \#7 & A list of the included primary studies on SA
    \\ 
    \bottomrule

    \end{tabular}

\end{table}

\subsection{Synthesis and analysis approach}\label{sec:synthesis-phase-1}

The extracted information was tabulated and visualized using graphs. For example, the total number of primary studies in each secondary study and the frequency of publications over time. We also visualized the timeline covered by the selected secondary studies. Furthermore, we identified an overlap between the primary studies of the selected secondary studies.

\subsection{Quality assessment}\label{sec:quality-assesment}
We adopted the quality assessment criteria provided by the York University Centre for Reviews and Dissemination (CRD) guide for reviews \cite{dissemination2009systematic} to assess the five selected secondary studies. To answer the five CRD/DARE questions, we used the criteria proposed by Budgen et al. \cite{budgen2020support}. The first author performed the quality assessment on the five selected secondary studies. 

\subsection{Validity threats} \label{sec:validity-threats}

In the following, we discuss potential validity threats to this review and our mitigation strategies.

\paragraph{Missing relevant secondary studies:} One possible threat to this review is the missing of relevant secondary studies. We mitigated this by carefully developing our search string based on the aim of the review, i.e., our search string searches for secondary studies that report a review of SA literature. In addition, we have used two indexing (i.e., Scopus and Web of Science) and two publisher-specific databases (i.e., IEEE Xplore and ACM) without restricting our search to papers from specific years to have sufficient coverage. These databases sufficiently cover SE literature \cite{kitchenham2016evidence,usman2021quality}.

\paragraph{Bias in paper selection:} Another threat to this review is bias during the paper selection process. We mitigate this risk by ensuring that every paper is reviewed by two reviewers. We also performed piloting before the actual study selection process. We had good agreements during the selection process, and there were only a few disagreements that we resolved through discussions and reconsideration (see details in Section \ref{sec:study-selection-phase-1}).

\section{Results and analysis}\label{phase-1-results}

This section presents the results of our research question: \textit{\textbf{RQ: What are the characteristics of secondary studies on SA?}}

\begin{figure}[!h]
    \centering
    \fbox{
    \begin{subfigure}[b]{0.45\textwidth}
        \includegraphics[width=\textwidth]{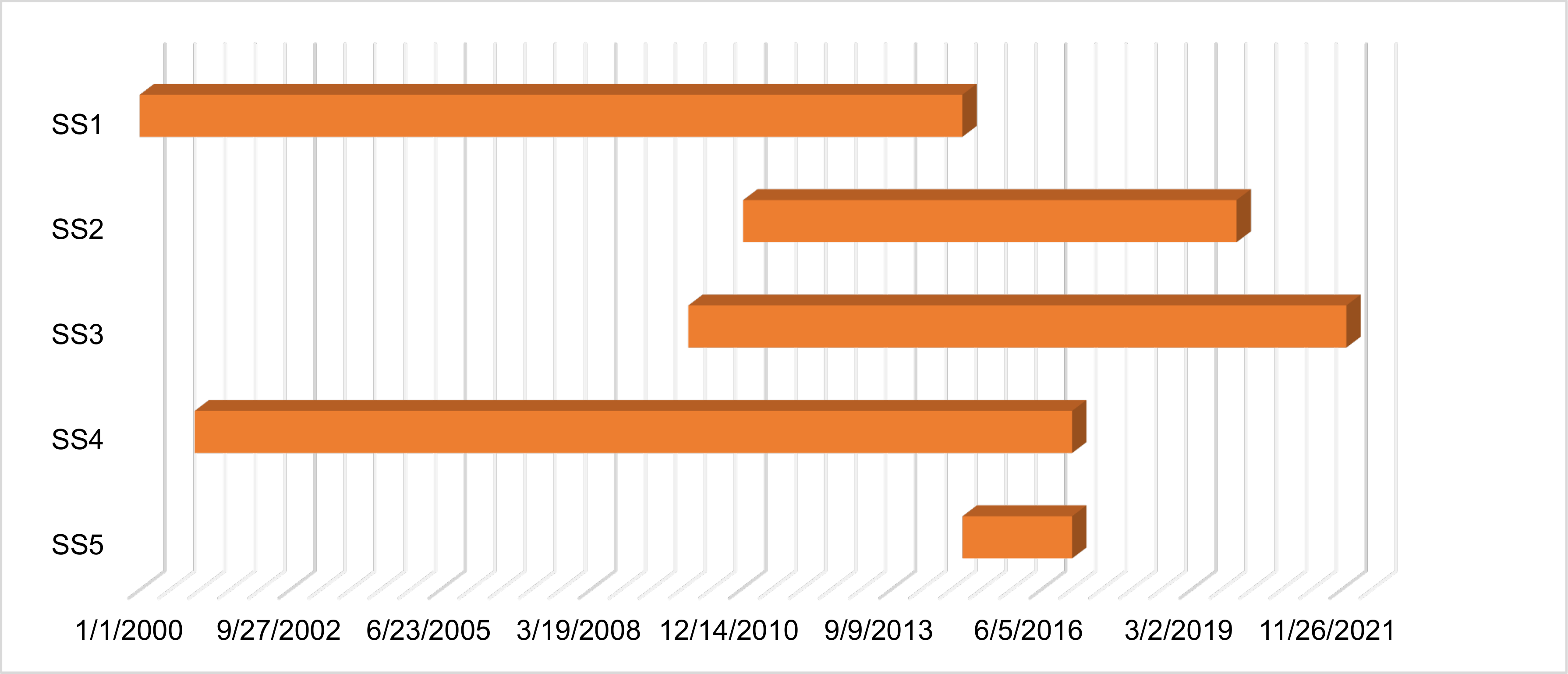}
        \caption{Timeline covered by secondary studies}
        \label{fig:timeline-of-SSs-phase-1}
    \end{subfigure}
    \quad
    \begin{subfigure}[b]{0.45\textwidth}
        \includegraphics[width=\textwidth]{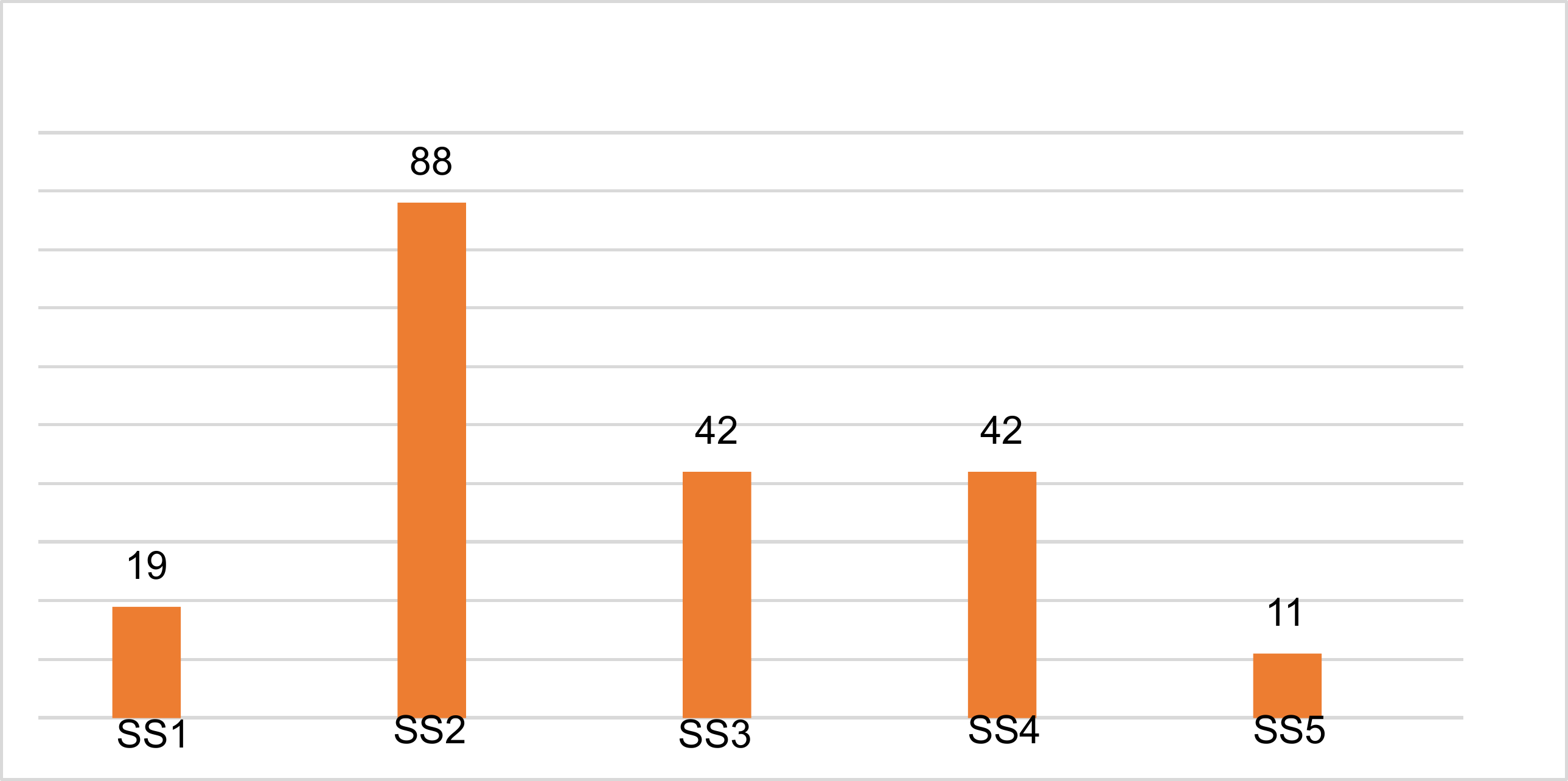}
        \caption{Count of primary studies in selected secondary studies}
        \label{fig:count-primary-studies-phase-1}
    \end{subfigure}
    }
    \caption{Timeline covered by secondary studies on SA with the count of their primary studies}
    \label{fig:search-timeline-and-count-phase-1}
\end{figure}

We selected five secondary studies (see details in Table \ref{tab:selected-SSs-phase-1}) following the selection criteria described in Section \ref{sec:study-selection-phase-1}.
In Figure \ref{fig:search-timeline-and-count-phase-1}, we show (a) the timeline of the primary research covered by these studies and (b) the count of the included primary studies in each secondary study.

\begin{table}[h]
    \centering
    \caption{Selected secondary studies on SA}
    \vspace{-2mm}
    \label{tab:selected-SSs-phase-1}
    \begin{tabular}{p{0.6cm}p{1.5cm}p{6cm}p{5.2cm}p{0.6cm}}
    \toprule
\textbf{Id} & \textbf{Pub. Year} & \textbf{Title}  & \textbf{Venue} & \textbf{Ref.} 
\\ 
\midrule

SS1 & 2015
& Software Analytics to Software Practice: A Systematic Literature Review & International Workshop on Big Data Software Engineering & \cite{abdellatif_software_2015}
\\    
\midrule

SS2 & 2021
& Big Data analytics in Agile software development: A systematic mapping study & Information and Software Technology & \cite{biesialska2021big}
\\ 
\midrule

SS3 & 2023
& Software Development Analytics in Practice: A Systematic Literature Review
& Archives of Computational Methods in Engineering
 & \cite{caldeira2023software}
\\ 
\midrule

SS4 & 2017
 & Software analytics for web usability: A systematic mapping
 & Computational Science and Its Applications 
 & \cite{gervasi_software_2017}
\\ 
\midrule

SS5 & 2017
 & Towards greener software engineering using software analytics: A systematic mapping
 & Euromicro Conference on Software Engineering and Advanced Applications
&\cite{anwar_towards_2017}
\\ 
\bottomrule

\end{tabular}

\end{table}

In Table \ref{tab:quality-assessment}, we show the summary of the quality assessment of the selected secondary studies.
Table \ref{tab:quality-assessment} shows that all studies have reported inclusion/exclusion criteria (C1). These studies have provided sufficient details of the selected primary studies (C5) and have used, to some extent, an adequate search strategy (C2). Both systematic literature reviews have assessed the quality of the included primary studies (C4).

Among the included studies, except for C3, all other criteria are adequately covered in most of the secondary studies, particularly C1, C2, and C5. This indicates that the identified literature in these studies could be sufficient to provide a map of the existing literature on the topic, provided that a sufficient timeline is covered in these studies. Furthermore, it is evident that there is generally a lack of synthesis of the evidence in these studies.

\begin{table}[!h]
    \centering
    \scriptsize
    \caption{Quality assessment of the identified secondary studies}
    \vspace{-2mm}
    \label{tab:quality-assessment}
    \begin{tabular}{p{9.5cm}p{1cm}p{1cm}p{0.9cm}p{0.9cm}p{0.5cm}}
    \toprule
     \textbf{Criterion and interpretation, Yes (Y), Partly (P), and No (N)} & \multicolumn{5}{p{5cm}}{\textbf{Studies score (Y: 1; P: 0.5; N: 0)}} 
    \\   
     & SS1*** & SS2** & SS3*** & SS4** & SS5** \\ \midrule
    \textbf{C1:} ``Were inclusion/exclusion criteria reported?'' \newline \textbf{Y:} The criteria are clearly defined. \newline \textbf{P:} The criteria are implicitly defined. \newline \textbf{N:} The criteria are not defined and can not be inferred readily.
    & 1 & 1 & 1 & 1 & 1 \\ \midrule
    
    \textbf{C2:} ``Was the search adequate?'' \newline \textbf{Y:} The authors have searched four or more digital libraries and included additional search strategies OR identified and referenced all journals addressing the topic of interest. \newline \textbf{P:} Searched three or four digital libraries with no extra search strategies OR searched a defined but restricted set of journals and conference proceedings. \newline \textbf{N:} Searched up to two digital libraries or an extremely restricted set of journals.
    & 0 & 1 & 0.5 & 0.5 & 0.5 \\ \midrule
    
    \textbf{C3:} ``Were the included studies synthesized?'' \\
    
    \textbf{C3a:} ``Was the evidence actually synthesized and aggregated, or merely summarized?'' \newline \textbf{Y:} The authors have performed a meta-analysis or used another form of synthesis for all the data of the study. \newline \textbf{P:} Synthesis has been performed for some of the data from some of the primary studies. \newline \textbf{N:} There is no real/explicit synthesis (as in a mapping study).
    & 0 & * & 0 & * & * \\ 
    
    \textbf{C3b:} ``Was the quality of individual studies taken into account in the synthesis?'' \newline \textbf{Y:} To some extent, yes. \newline \textbf{N:} No.
    & 0 & * & 0 & * & * \\ \midrule
    
    \textbf{C4:} ``Was the quality of the included studies assessed?'' \newline \textbf{Y:} The authors have explicitly defined quality criteria and extracted them from each primary study. \newline \textbf{P:} The research question involved quality issues that are addressed by the study. \newline \textbf{N:} There is no explicit quality assessment for individual studies.
    & 1 & * & 1 & * & * \\ \midrule
    
    \textbf{C5:} ``Are sufficient details about the individual included studies presented?'' \newline \textbf{Y:} Information for each study has been provided. \newline \textbf{P:} Information has been provided in a summarized form only. \newline \textbf{N:} The results of individual studies have not been specified.
    & 1 & 1 & 1 & 0.5 & 0.5 \\ \midrule

    \raggedleft \textbf{Total score =} & 3/6 & 3/3 & 3.5/6 & 2/3 & 2/3 \\ \bottomrule

\multicolumn{6}{p{15cm}}{\footnotesize{*We do not apply this criterion to a systematic mapping study. **Systematic mapping study. \newline ***Systematic literature review.}}

\end{tabular}
\end{table}

\subsection*{RQ1: How is SA defined?} \label{rq1-phase-1}

The selected secondary studies define SA as cited below:

\begin{itemize}

\item \textbf{SS1:} Abdellatif et al. \cite{abdellatif_software_2015}: \textit{``Software analytics (SA) represents a branch of big data
analytics. SA is concerned with the analysis of all software artifacts, not only source code. Its importance comes from the need to extract support insights and facts from the available software artifacts to facilitate decision making.''}

\item \textbf{SS2:} Biesialska et al. \cite{biesialska2021big} did not explicitly define SA in their paper. However, they relate it to big data and data analytics, i.e., \textit{``Software analytics or data analytics belong to a common category. In order to provide non-obvious insights and improve the decision-making process, they need to employ ML, DL, or AI methods in some shape or form. Those methods require a significant amount of data (often complex or unstructured) to train models.''}

\item \textbf{SS3:} Caldeira et al. \cite{caldeira2023software} provide two definitions of SA: (i)\textit{``The term `software analytics' (SA) emerged naturally expressing the work of several research groups aiming to expand the traditional scope on analyzing software artifacts by means of mining software repositories. These groups conducted cutting-edge research and technology innovation in an interdisciplinary area that spans big data, machine learning, systems, and software engineering. This approach led software practitioners to perform data exploration and analysis in order to obtain insightful and actionable information for completing various tasks around software systems, software users, and software development processes.''}, and (ii) \textit{``Software Development Analytics (SDA), the adoption
of analytics methods with the focus on the management of
software development projects."}

\item \textbf{SS4:} Pellizon et al. \cite{gervasi_software_2017}: \textit{``The software analytics area is concerned with collecting, exploring and analyzing this large amount of data aiming to transform it into actions to improve the product and its usability.''
}
\item \textbf{SS5:} Anwar et al. \cite{anwar_towards_2017} did not explicitly describe SA in their paper; instead, they define the usefulness of SA: \textit{``Software analytics could support practitioners with actionable and timely information as it combines information from different software artifacts and converts it into useful information.''}

\end{itemize}

Table \ref{tab:SA-definitions-analysis-phase-1} shows an analysis of the definitions of SA in the selected secondary studies. The diversity in defining SA makes defining the boundaries of what we should consider SA research problematic.

In Table \ref{tab:SA-definitions-analysis-phase-1}, we show the main themes we noticed in the statements describing SA and its purpose in the included secondary studies. Below, we analyze these further using the six interrogative words as proposed by Petersen et al. \cite{Petersen2024}. 

\textbf{Why is a solution required?} To facilitate decision-making regarding the software processes, products, or their usability 

\textbf{What is the solution?} Big data analytics, AI, ML, or DL to derive actionable and non-obvious insights.

\textbf{How does the solution work?} By using any available large data (structured or unstructured), or various other software artifacts in software repositories.

\textbf{Where in the process is the solution required?} Entire software lifecycle.

\textbf{Who are the users of the solution?} Various practitioners. Here, we see that SA is primarily focused on software practitioners and not researchers. This is again obvious in the following aspect of timing.

\textbf{When refers to timing.} Timely information, i.e., the information should be readily available when relevant decisions in the software development life-cycle arise. So, the feedback time for the solution is bound by the frequency and urgency of the practical decisions.

Based on the above analysis of the definitions of SA in the selected secondary studies, we define SA as follows:

\vspace{\baselineskip}
\noindent
\fbox{\begin{minipage}[!]{0.98\columnwidth}%
\textit{``Software Analytics is concerned with \textbf{collecting, exploring and analyzing} a \textbf{large amount of data and software artifacts} using AI (e.g., ML and DL) and other big data and data analytics techniques to support \textbf{software practitioners} in \textbf{decision making} related to software development by providing\textbf{ non-obvious, timely, insights and actionable information}.''}
\end{minipage}}

\begin{table}[!ht]
\centering
\small
\caption{Analysis of the definitions of SA used in the selected SSs}
\vspace{-2mm}
\label{tab:SA-definitions-analysis-phase-1}
\begin{tabular}{p{0.6cm}p{4cm}p{5cm}p{2.4cm}p{2.5cm}}
\\ \toprule
  \textbf{Id} & \textbf{Technology/Methods}  & \textbf{Aims/Outcomes} & \textbf{Data sources} & \textbf{Stakeholders}
   \\ \midrule
SS1 & Big data & Facilitate decision-making, provide insights and facts &  Available software artifacts & Practitioners*
   \\ \midrule

SS2 & Big data, data analytics, artificial intelligence (AI), machine learning (ML), deep learning (DL) & Improve decision-making, provide non-obvious insights &  Large data, complex data, unstructured data & Practitioners*
   \\ \midrule

SS3 &  Mining software repositories, big data, ML & Enable data exploration and data analysis, provide insightful and actionable information,  management of software development projects &  Software artifacts &  Practitioners
   \\ \midrule
   
SS4 & -- & Actions to improve product and its usability &  Large amount of data & Practitioners*
   \\ \midrule

SS5 & -- & Actionable information, timely information, useful information  & Use different software artifacts & Practitioners
   \\ \bottomrule   

\multicolumn{5}{l}{\footnotesize{* Indicates that the practitioner as a target stakeholder is implied in the definition.}}

\end{tabular}
\end{table}

\subsection*{RQ2: Which topics have been studied?}\label{Rq2-phase-1}

We found that the selected secondary studies have targeted different areas of SE when identifying the usage of SA in SE (see overview in Table \ref{tab:selected-SSs-phase-1}). Two secondary studies generally focused on aggregating research on SA for SE [SS1, SS3]. The remaining three focused on SA in different areas of SE, that is, exploring SA usage for web usability [SS4], agile software development [SS2], and green SE [SS5]. 

\subsection*{RQ3: What search strategies have been used?}\label{Rq3-phase-1}

Table \ref{tab:search-of-SSs-phase-1} shows the search used in each selected secondary study. The search strings in SS1 and SS3 are targeted at work that is explicitly SA-based. However, SS3 is also limited to studies that use techniques such as data mining and big data and research methods such as in vivo and empirical studies. SS2 used a manual search strategy in selected venues and focused on SA for agile software development. SS4 focuses on web usability, and SS5 focuses on green SE.

\begin{table}[ht]
    \centering
    \caption{Search string/strategy used in the selected secondary studies}
    \vspace{-2mm}
    \label{tab:search-of-SSs-phase-1}
    \begin{tabular}{lp{15cm}}
    \toprule
\textbf{Id} & \textbf{Search string/strategy}
\\ \midrule 

SS1 & \textit{``Software analytics'' OR ``Software analytic'' OR ``Software development analytics'' OR ``Software development analytic''}
\\ \midrule

SS2 & \textit{Manual search in a pre-selected list of venues (e.g., Information and Software Technology Journal, International Conference on Software Engineering, and Software Quality Journal), and backward and forward snowballing}
\\ \midrule

SS3 & \textit{(``software analytics'' OR ``software development analytics'') AND (``process mining'' OR ``data mining'' OR ``big data'' OR ``data science'') AND (``study'' OR ``empirical'' OR ``evidence based'' OR ``experimental'' OR ``in vivo'')}
\\  \midrule

SS4 & \textit{(capture OR collect OR analyze) AND (user behavior OR usability OR user experience) AND (web)}
\\ \midrule

SS5 & \textit{(``Software engineering'') AND (green* OR sustain* OR ``energy consumption'' OR ``energy-aware'' ``energy-efficient'' OR ``resource optimization'') AND NOT (feedback OR domestic OR electric* OR mechanic* OR industrial OR chemical* OR bio* OR hardware OR net*)}
\\  \bottomrule

\end{tabular}
\end{table}

Table \ref{tab:targeted-dbs-phase-1} shows the targeted databases in each selected secondary study. The publisher-specific databases IEEE Xplore and ACM seem to be commonly used and have been targeted in 3 out of 5 studies with other databases, such as Scopus and Web of Science. SS1 has only used IEEE Xplore and ACM, which could be considered a limitation of their search coverage. These two databases do not provide full coverage of SE literature. Systematic literature review guidelines \cite{kitchenham2013systematic, dyba2007applying} in SE also recommend using at least one indexing database, such as Scopus, along with publisher-specific databases.

\begin{table}[!h]
    \centering
    \caption{Targeted databases}
    \vspace{-2mm}
    \label{tab:targeted-dbs-phase-1}
    \begin{tabular}{lccccc}
    \toprule
\textbf{Databases} & \textbf{SS1} & \textbf{SS2} & \textbf{SS3} & \textbf{SS4} & \textbf{SS5}
\\ 
\midrule
IEEE Xplore & \checkmark & & \checkmark & &  \checkmark  \\  \midrule
ACM  & \checkmark & & \checkmark & &    \checkmark \\   \midrule

Scopus & & & \checkmark & \checkmark &    \\   \midrule

Web of Science & & & \checkmark & &  \checkmark \\   \midrule

Google Scholar & & & \checkmark & &  \\   \midrule

Wiley Online & & & \checkmark & &  \checkmark \\   \midrule

ScienceDirect & & & \checkmark & &  \checkmark \\   \midrule

SpringerLink  & & & \checkmark & &  \checkmark \\    \midrule

Selected Venues & & \checkmark & & &  \\   \bottomrule
    
\end{tabular}
\end{table}

\subsection*{RQ4: What is the overlap of primary studies?}\label{Rq4-phase-1}

In the selected secondary studies, we found only two common papers between SS1 and SS2, one between SS1 and SS3, and no other overlaps (see Table \ref{tab:overlaps_and_cocitations} for an overview).
The small overlaps between existing secondary studies could be the result of insufficient/different search strategies and databases used in these secondary studies (see details of the search strategy used in these studies in Section \ref{Rq3-phase-1}).

\begin{table}
    \centering
    \caption{Overlaps\textbf{**} and co-citations (colored cells) among included secondary studies (indicating awareness of related secondary studies)}
    \vspace{-2mm}
    \label{tab:overlaps_and_cocitations}
    \begin{tabular}{|c|c|c|c|c|c|}
        \hline
   & SS1 (2015) & SS2 (2021) & SS3 (2023) & SS4 (2017) & SS5 (2017) \\ \hline
   SS1 (2015)   &  NA & \cellcolor{olive!25}1.9\% & \cellcolor{olive!25}1.66\% & 0 & 0 \\ \hline
   SS2 (2021)   &  \cellcolor{olive!25}1.9\% & NA & 0 & 0 & 0 \\ \hline
   SS3 (2023)   &  \cellcolor{olive!25}1.66\% & 0 & NA & 0 & 0 \\ \hline
   SS4 (2017)   & 0 & 0 & 0 & NA & 0 \\ \hline
   SS5 (2017)   & 0 & 0 & 0 & 0 & NA \\ \hline
   \multicolumn{6}{|p{13cm}|}{\textbf{Total overlap =  0.4\%.}}
   \\ \hline
   \multicolumn{6}{p{13cm}}{\footnotesize \textbf{**}The corrected covered
area, a single number indicating the total overlap of PSs for all included secondary studies \cite{borstler2023double}. Total overlap =  0.4\%, which is very low. Up to $<$6\%, the overlap is considered as ``slight.' 0 = no overlap and NA = not applicable.}
   
    \end{tabular}
\end{table}

Furthermore, as shown in Table \ref{tab:overlaps_and_cocitations}, among these studies, only SS2 (2021) and SS3 (2023) have cited SS1 (2015). SS2 reported that none of the previously published related studies extensively cover the topic of interest by addressing research questions similar to ours. SS3 reported that their goal is to expand the existing knowledge on SA by adapting and extending the data perspectives, dimensions, and concerns identified and used in the related work.

\section{Discussion} \label{sec:discussion}

Below, we discuss the findings of this review based on the five selected secondary studies on the usage of SA in SE.

The timeline covered by the primary studies included in the five selected secondary studies ranges from 2000 to 2021. Despite the overlapping timelines of the selected secondary studies (see Figure \ref{fig:timeline-of-SSs-phase-1}), there is negligible overlap of primary studies between the secondary studies. Thus, each of these studies provides an isolated view, and together they provide a fragmented view, i.e., there is no ``common picture'' of the area as these studies have used distinct classification schemes. 
For example, SS1 and SS3 classify the included primary studies in the following categories (C) or areas. \textbf{SS1} -- C1: Maintainability and reverse engineering; C2: Team collaboration and dashboard; C3: Incident management and defect prediction; C4: Software analytics platform; C5: Software effort estimation. \textbf{SS3 --} C1: Implementation; C2: Maintenance; C3: Testing; C4: Debugging; C5: Operations.
Thus, there is a need to structure the existing work using a single classification scheme in a consistent map.

One possible explanation for the lack of overlap of primary studies between these secondary studies could be the use of the different search strategies used in these studies. For example, SS1 and SS3 focused on generally aggregating research on SA for SE. However, their search string differs. SS1 search string searches for literature that uses the terms ``software analytics'' or ``software development analytics.'' SS3 has these terms with additional criteria that include terms such as ``big data'' and ``data mining'' and type of study, e.g., ``empirical'' and ``evidence-based''. Similarly, the search strategies for SS2, SS4, and SS5 have their own criteria.
 
Considering the different areas, the number of included primary studies, and the timeline covered by the selected secondary studies, an overview study combining the primary studies included in these secondary studies could be adequate to provide a comprehensive overview of the usage of SA in SE.   

In summary, the above-mentioned findings indicate that the existing secondary studies on the usage of SA in SE do not provide a comprehensive and consistent map, which has the following limitations: (a) for researchers, it limits the possibilities of identifying future research directions, while (b) for practitioners, there is no extensive collection of what potentially relevant SA solutions exist. Consequently, there is a need for an overview study that combines the primary studies included in these secondary studies to provide a consistent and comprehensive map of the usage of SA in SE.

\section{Conclusion and future work} \label{sec:conclusion-future-work}

In this study, we analyzed existing secondary studies on the usage of SA for SE. We found five secondary studies on the topic covering the primary research from 2000 to 2021. Despite the overlapping objectives and search timeframes of these secondary studies, there is a negligible overlap of primary studies between these secondary studies. This has resulted in a fragmented view of the literature on the topic. In addition, these studies have used distinct classification schemes. Thus, there is a need for an overview study that combines these studies to provide a more comprehensive overview of the topic using a consistent mapping scheme.

In the future, we aim to conduct a comprehensive overview study to provide a map of the existing literature on SA for SE by collectively analyzing primary studies from the identified secondary studies.

\section*{Acknowledgments}
This work has been supported by ELLIIT; the Strategic Research Area within IT and Mobile Communications, funded by the Swedish Government. The work has also been supported by a research grant for the GIST project (reference number 20220235) from the Knowledge Foundation in Sweden.

\bibliographystyle{acm}
\bibliography{main}

\begin{thebibliography}{10}

\bibitem{abdellatif_software_2015}
{\sc Abdellatif, T.~M., Capretz, L.~F., and Ho, D.}
\newblock Software analytics to software practice: A systematic literature review.
\newblock In {\em 2015 {IEEE}/{ACM} 1st International Workshop on Big Data Software Engineering}, {IEEE}, pp.~30--36.

\bibitem{anwar_towards_2017}
{\sc Anwar, H., and Pfahl, D.}
\newblock Towards greener software engineering using software analytics: A systematic mapping.
\newblock In {\em 2017 43rd Euromicro Conference on Software Engineering and Advanced Applications ({SEAA})}, pp.~157--166.

\bibitem{biesialska2021big}
{\sc Biesialska, K., Franch, X., and Munt{\'e}s-Mulero, V.}
\newblock Big data analytics in agile software development: A systematic mapping study.
\newblock {\em Information and Software Technology 132\/} (2021), 106448.

\bibitem{BorstlerAP23}
{\sc B{\"{o}}rstler, J., Ali, N.~B., and Petersen, K.}
\newblock Double-counting in software engineering tertiary studies - an overlooked threat to validity.
\newblock {\em Inf. Softw. Technol. 158\/} (2023), 107174.

\bibitem{borstler2023double}
{\sc B{\"o}rstler, J., bin Ali, N., and Petersen, K.}
\newblock Double-counting in software engineering tertiary studies—an overlooked threat to validity.
\newblock {\em Information and Software Technology 158\/} (2023), 107174.

\bibitem{budgen2020support}
{\sc Budgen, D., Brereton, P., Williams, N., and Drummond, S.}
\newblock What support do systematic reviews provide for evidence-informed teaching about software engineering practice?
\newblock {\em e-informatica software engineering journal 14}, 1 (2020), 7--60.

\bibitem{caldeira2023software}
{\sc Caldeira, J., Brito~e Abreu, F., Cardoso, J., Sim{\~o}es, R., Oliveira, T., and Pereira~dos Reis, J.}
\newblock Software development analytics in practice: A systematic literature review.
\newblock {\em Archives of Computational Methods in Engineering\/} (2023), 1--40.

\bibitem{dissemination2009systematic}
{\sc Dissemination, C.}
\newblock Systematic reviews: Crd’s guidance for undertaking reviews in healthcare.
\newblock {\em York: University of York NHS Centre for Reviews \& Dissemination\/} (2009).

\bibitem{dyba2007applying}
{\sc Dyba, T., Dingsoyr, T., and Hanssen, G.~K.}
\newblock Applying systematic reviews to diverse study types: An experience report.
\newblock In {\em First international symposium on empirical software engineering and measurement (ESEM 2007)\/} (2007), IEEE, pp.~225--234.

\bibitem{garousi2016systematic}
{\sc Garousi, V., and M{\"a}ntyl{\"a}, M.~V.}
\newblock A systematic literature review of literature reviews in software testing.
\newblock {\em Information and Software Technology 80\/} (2016), 195--216.

\bibitem{IftikharABU24}
{\sc Iftikhar, U., Ali, N.~B., B{\"{o}}rstler, J., and Usman, M.}
\newblock A tertiary study on links between source code metrics and external quality attributes.
\newblock {\em Inf. Softw. Technol. 165\/} (2024), 107348.

\bibitem{kitchenham2013systematic}
{\sc Kitchenham, B., and Brereton, P.}
\newblock A systematic review of systematic review process research in software engineering.
\newblock {\em Information and software technology 55}, 12 (2013), 2049--2075.

\bibitem{kitchenham2015evidence}
{\sc Kitchenham, B.~A., Budgen, D., and Brereton, P.}
\newblock {\em Evidence-based software engineering and systematic reviews}, vol.~4.
\newblock CRC press, 2015.

\bibitem{kitchenham2016evidence}
{\sc Kitchenham, B.~A., Budgen, D., and Brereton, P.}
\newblock {\em Evidence-based software engineering and systematic reviews}.
\newblock CRC press, 2016.

\bibitem{laiq2024industrial}
{\sc Laiq, M., Ali, N.~b., B{\"o}rstler, J., and Engstr{\"o}m, E.}
\newblock Industrial adoption of machine learning techniques for early identification of invalid bug reports.
\newblock {\em Empirical Software Engineering 29}, 5 (2024), 130.

\bibitem{marijan2013test}
{\sc Marijan, D., Gotlieb, A., and Sen, S.}
\newblock Test case prioritization for continuous regression testing: An industrial case study.
\newblock In {\em 2013 IEEE International Conference on Software Maintenance\/} (2013), IEEE, pp.~540--543.

\bibitem{nurdiani2016impacts}
{\sc Nurdiani, I., B{\"o}rstler, J., and Fricker, S.~A.}
\newblock The impacts of agile and lean practices on project constraints: A tertiary study.
\newblock {\em Journal of Systems and Software 119\/} (2016), 162--183.

\bibitem{gervasi_software_2017}
{\sc Pellizon, L.~H., Choma, J., da~Silva, T.~S., Guerra, E., and Zaina, L.}
\newblock Software analytics for web usability: A systematic mapping.
\newblock In {\em Computational Science and Its Applications – {ICCSA} 2017}, O.~Gervasi, B.~Murgante, S.~Misra, G.~Borruso, C.~M. Torre, A.~M.~A. Rocha, D.~Taniar, B.~O. Apduhan, E.~Stankova, and A.~Cuzzocrea, Eds., vol.~10409. Springer International Publishing, pp.~246--261.
\newblock Series Title: Lecture Notes in Computer Science.

\bibitem{Petersen2024}
{\sc Petersen, K., B\"{o}rstler, J., Ali, N.~B., and Engstr\"{o}m, E.}
\newblock Revisiting the construct and assessment of industrial relevance in software engineering research.
\newblock In {\em Proceedings of the 1st IEEE/ACM International Workshop on Methodological Issues with Empirical Studies in Software Engineering\/} (New York, NY, USA, 2024), WSESE '24, Association for Computing Machinery, p.~17–20.

\bibitem{pillai2017consolidating}
{\sc Pillai, S.~P., Madhukumar, S., and Radharamanan, T.}
\newblock Consolidating evidence based studies in software cost/effort estimation—a tertiary study.
\newblock In {\em TENCON 2017-2017 IEEE Region 10 Conference\/} (2017), IEEE, pp.~833--838.

\bibitem{rakha2018revisiting}
{\sc Rakha, M.~S., Bezemer, C.-P., and Hassan, A.~E.}
\newblock Revisiting the performance of automated approaches for the retrieval of duplicate reports in issue tracking systems that perform just-in-time duplicate retrieval.
\newblock {\em Empirical Software Engineering 23\/} (2018), 2597--2621.

\bibitem{TranUBA21}
{\sc Tran, H. K.~V., Unterkalmsteiner, M., B{\"{o}}rstler, J., and Ali, N.~B.}
\newblock Assessing test artifact quality - {A} tertiary study.
\newblock {\em Inf. Softw. Technol. 139\/} (2021), 106620.

\bibitem{usman2021quality}
{\sc Usman, M., Ali, N.~B., and Wohlin, C.}
\newblock A quality assessment instrument for systematic literature reviews in software engineering.
\newblock {\em arXiv preprint arXiv:2109.10134\/} (2021).

\bibitem{zhang2013sainpractice}
{\sc Zhang, D., Han, S., Dang, Y., Lou, J.-G., Zhang, H., and Xie, T.}
\newblock Software analytics in practice.
\newblock {\em IEEE software 30}, 5 (2013), 30--37.

\bibitem{zhou2016map}
{\sc Zhou, X., Jin, Y., Zhang, H., Li, S., and Huang, X.}
\newblock A map of threats to validity of systematic literature reviews in software engineering.
\newblock In {\em 2016 23rd Asia-Pacific Software Engineering Conference (APSEC)\/} (2016), IEEE, pp.~153--160.

\end{thebibliography}

\end{document}